\documentclass{aa}
\usepackage{graphicx}
\usepackage{epsfig}
\def\arcsec{\hbox{$^{\prime\prime}$}}
\def\degree{\hbox{$^{\circ}$}}

\begin{document}
\title{
Near-IR spectroscopy of planetary nebulae precursors
\thanks{Based on observations collected at the European Southern Observatory
(La Silla, Chile)}}
\author{D.A. Garc\'\i a-Hern\'andez \inst{1}
\and A. Manchado\inst{1,2}
\and P. Garc\'\i a-Lario \inst{3}
\and C. Dom\'\i nguez-Tagle\inst{1}
\and G.M. Conway \inst{4}
\and F. Prada \inst{5}}
\institute{Instituto de Astrof\'{\i}sica de Canarias, E-38205 La Laguna, 
Tenerife, Spain\\
\email{(agarcia, amt, carlosd)@ll.iac.es}
\and Consejo Superior de Investigaciones Cient\'{\i}ficas, CSIC
\and ISO Data Centre, Science Operations and Data Systems Division, 
Research and Scientific Support Department of ESA, VILSPA, E-28080 Madrid,
 Spain\\
 \email{pgarcia@iso.vilspa.esa.es}
\and Department of Physics and Astronomy. University of Calgary, 
Alberta T2N 1N4, Canada 
\and Centro Astron\'omico Hispano$-$Alem\'an. Apartado de Correos 511, E-04080,
Almer\'\i a, Spain}
\offprints{D.A. Garc\'\i a-Hern\'andez}
\date{Received July 31, 2001; accepted March 18, 2002}
\authorrunning{Garc\'\i a-Hern\'andez et al.}
\titlerunning{Near$-$IR spectroscopy of planetary nebulae precursors}

\abstract{
We present near-IR spectroscopy  of a sample of 30
IRAS sources recently identified as late AGB stars, post-AGB stars or early
 PNe. The spectra obtained are  
centered at various wavelengths covering the molecular
hydrogen $v$=1$\rightarrow$0 S(1) 2.122 $\mu$m and  $v$=2$\rightarrow$1 
S(1) 2.248 $\mu$m emission lines, the  recombination lines of hydrogen 
Br$\gamma$ 2.166 $\mu$m,  Pf$\gamma$ 3.741 $\mu$m and  Br$\alpha$ 
4.052 $\mu$m, and the CO[$v$=2$\rightarrow$0] first overtone  
bandhead at 2.294 $\mu$m. As a result of these observations we have 
detected molecular hydrogen emission for the first time in 9 of these
sources  and confirmed a previous detection by Weintraub et al. 
(1998). This increases from 4 to 13  the total number of proto-PNe detected 
in molecular hydrogen. In most cases, the positive detections  
also show emission in the recombination lines of hydrogen (with the exception
of IRAS 17150$-$3224) indicating that the onset of molecular hydrogen 
emission takes place in the post-AGB phase, very shortly before the 
nebula becomes ionized. When the molecular 
hydrogen is fluorescence-excited the detection rate is found to 
be directly correlated with the evolutionary stage of the central star,
rather than with the nebular morphology. When the temperature of the 
central star is hot enough, fluorescence excitation can be
induced by the absorption of UV photons escaping from the  
rapidly evolving central post-AGB star.  In contrast, shocked-excited 
molecular hydrogen is detected only in strongly bipolar proto-PNe, 
sometimes even at an early stage in the post-AGB phase. Shock-excitation 
is the consequence 
of the interaction of the fast post-AGB wind with the slow wind material 
ejected during the AGB. The strong correlation of shocked-excited molecular
hydrogen emission with bipolarity found  
confirms the result previously reported by Kastner et al. (1996) in evolved 
PNe. However, our results show that this correlation does not 
exist in the case of fluorescence-excited molecular hydrogen. 
}
\maketitle

\section{Introduction}

The short transition phase between the end of the asymptotic giant branch
(AGB)  and the formation of a new planetary nebula (PN) is still poorly
understood.  The main reason for this is probably twofold. First,
only a small number of objects are known in this phase, as this is a 
short-lived evolutionary stage. Second,  in many cases, stars
evolve through this phase heavily obscured by the strong
circumstellar envelopes formed during the previous AGB phase, thus making
observations difficult. 

 However, this is an evolutionary stage when crucial changes occur that 
completely determine the subsequent evolution of the individual star as a PN. 
Eventually, this also results in changes in the
 overall chemical evolution of the Galaxy,  since
 huge amounts of processed material are transferred to the 
interstellar medium as a consequence of the strong mass loss experienced 
at the very end of the AGB phase. 

 It is also during this short transition phase when
the ionization of the gas ejected during the previous AGB phase 
takes place. However, it is now well established that in most cases only 
a small fraction of it  gets eventually ionized in comparison
with the total amount of mass expelled  by the central star 
during the AGB phase (tenths of a solar mass of ionized gas 
compared to up to several solar masses ejected during the AGB
phase in the case of the more massive progenitors) (Pottasch 1980, 
1992; Kwok 1994). 
Most of this material remains neutral  in the form of dust 
grains, molecules or atoms which can be detected easily at infrared or
radio wavelengths even when the star has completely evolved and turned into 
a PN. 

 From the analysis of IRAS data, and especially 
with the advent of the higher quality ISO data, it has 
been possible to confirm the presence of large amounts of dust grains 
which emit thermally in the mid- and far-infrared in objects evolving from
the AGB to the PN stage. Through the analysis of 
the features observed in the ISO spectra it is possible to
determine the dominant gas and dust chemistry (O-rich or C-rich) 
in the stellar photosphere and in the circumstellar
envelope (Waters et al. 1999; Sylvester et al. 1999; Hrivnak et al. 2000).  

 The detection of molecules in the envelopes around PNe 
has also been successful at submilimeter and radio wavelengths over 
the last two decades.
Strong CO molecular emission has been detected in a large number of C-rich 
PNe (Huggins \& Healy 1989; Huggins et al. 1996) 
and  is also detectable around both C-rich and O-rich post-AGB stars 
(Likkel et al. 1991; Loup et al. 1993;
Neri et al. 1998, and references therein). Other molecules, like HCN or CN
are detected only in envelopes dominated by a 
C-rich chemistry (Lindqvist et al. 2000), while
in the case of circumstellar envelopes 
dominated by an O-rich chemistry strong molecular maser emission from 
SiO, H$_2$O and OH is usually detected when the star is in the AGB phase
(Engels 1979; Engels \& Lewis 1996; Nyman et al. 1998; te Lintel
 Hekkert 1991a; te Lintel Hekkert 
et al. 1991b). In many cases, the OH maser emission remains detectable 
beyond the AGB 
throughout the whole post-AGB evolution (te Lintel Hekkert \& Chapman 1996) 
and sometimes even  when the object has already become an ionized PN 
(the so-called OHPNe - Zijlstra et al. 1989; 1991). 
More complex molecules are difficult to detect, but in the circumstellar
envelopes of a few objects like
IRC+10216, AFGL 618 and AFGL 2688 more than 50 different molecular species 
have been identified in the last few years,
and the number and complexity of the new molecules found is continuously 
increasing (Wallerstein \& Knapp 1998;
Bujarrabal et al. 1988; Cernicharo et al. 1999).

 Since hydrogen is the main constituent of stellar  atmospheres one would
expect to find also a significant amount of neutral material locked up in the
form of molecular hydrogen around stars evolving from the AGB phase to the PN
stage (both O-rich and C-rich).  Following the first detection of molecular
hydrogen in the PN NGC 7027  (Treffers et al. 1976), more than 70 other PNe
have been detected in the last  few years by Webster et al. (1988), Zuckerman
\& Gatley (1988), Kastner et  al. (1996), Latter et al. (1995), Hora et al.
(1999) and Guerrero et al. (2000). As a  general trend, it is found that
molecular hydrogen emission is  predominantly detected in  type I bipolar PNe,
which are expected  to be the result of the evolution of the more massive AGB
progenitors.  It remains unclear, however, whether this is just an
observational selection effect.

Very few attempts have been made so far  
to search for molecular hydrogen around post-AGB stars or proto-PNe, the 
precursors of PNe. Molecular hydrogen emission was first detected  towards
 AFGL 618 by Beckwith et al. (1978) and 
AFGL 2688 by Gatley et al. (1988). More recently, 
Weintraub et al. (1998) looked at a small sample of five  
proto-PNe and detected molecular hydrogen emission in two of them (IRAS 
17441-2411 and AFGL 6815S=IRAS 17150-3224), bringing
to four the total number of bipolar proto-PNe with this type of emission. 
The  discovery and study of new transition objects showing
this emission at such an early stage 
is crucial for determining the mechanism of 
excitation of molecular hydrogen and the conditions needed to preserve this 
neutral material from destruction in more evolved PNe.

 This paper extends this search to a larger sample of transition 
objects, including late-AGB stars,  
post-AGB stars with various spectral types, as well as the 
youngest PNe known,  in order to look for possible correlations between the
onset of molecular hydrogen emission and the evolutionary stage and/or the 
morphology of the sources observed. 

 In Sect. 2 we explain the criteria followed to select the sources in the
sample. The spectroscopic observations made in the near infrared are described
in Sect. 3, while the results obtained are presented in Sect. 4 and 
discussed in Sect. 5. The conclusions derived from these results are given
in Sect. 6.

\section{Selection of the sample}

\begin{table*}[t!]
\caption{IRAS sources included in the survey and main observational properties.}
\begin{tabular}{lllll}
\hline
IRAS name& Other names&  Spectral Type& H$_{\alpha}$ emission& 
Size/Morphology\\
\hline
05284+1945  &  & OH/IR&     --& No optical counterpart\\
05341+0852  &  &  F4Iab& no&    $1.1\arcsec\times0.8\arcsec$ / Round-Elliptical\\
06530$-$0213&  &   F0I&    no&  $2.3\arcsec\times1.0\arcsec$ / Bipolar \\
06556+1623  &HD 51585, MWC 162& B1I& yes (strong)& Unknown\\
07027$-$7934&                 &   PN (WC11)& yes (strong)& $13\arcsec\times15\arcsec$ / Round-Elliptical\\
07134+1005  &SAO 96709, HD 56126&  F7I&  yes (weak)&$4.7\arcsec\times3.9\arcsec$ / Round-Elliptical\\
07331+0021  &AI CMi, RAFGL 5236& K3-5I& yes (weak)&Unknown\\
08005$-$2356 & & F3I& yes (strong)& $2.6\arcsec\times1.2\arcsec$ / Bipolar\\
08187$-$1905 &HD 70379& F7I& yes (weak)& Unknown\\
08242$-$3828 &       & OH/IR& --& No optical counterpart\\
08425$-$5116 &       & OH/IR& --& No optical counterpart\\
09024$-$5019 &       & ---& --& No optical counterpart\\
10178$-$5958 &Hen 3$-$401, C 1662& B1I?& yes (strong)& $2.0\arcsec\times14.5\arcsec$ / Bipolar\\
10197$-$5750 &Roberts 22, RAFGL 4104& A2I& yes (strong)& $7.8\arcsec\times4.5\arcsec$ / Bipolar\\
10215$-$5916 &RAFGL 4106& G1I& yes (weak)&  $\approx7\arcsec\times7\arcsec$ / Round-Elliptical \\
11438$-$6330 & & OH/IR& --& No optical counterpart \\
12067$-$4508 &RU Cen, SAO 223245& G2w& yes (weak)& Unknown\\
12175$-$5338 &SAO 239853& F2I& yes (weak)& Unknown\\
13421$-$6125 && OH/IR& --& No optical counterpart\\
14104$-$5819 && OH/IR& --& No optical counterpart\\
14331$-$6435 &Hen 3$-$1013& B8I& yes (strong)&  Unknown\\
16342$-$3814 & & M4& no& $2.6\arcsec\times0.7\arcsec$ / Bipolar\\
16594$-$4656 & & B7?& yes (strong)& $5\arcsec\times11\arcsec$ / Multipolar\\
17119$-$5926 &Hen 3$-$1357, Stingray Nebula& PN (O8V)& yes (strong)& $1.6\arcsec\times1.4\arcsec$ / Bipolar\\
17149$-$3053 & & M4?& no& Unknown\\
17150$-$3224 &RAFGL 6815S& G2I& no& $11\arcsec\times5\arcsec$ / Bipolar\\
17311$-$4924 &Hen 3$-$1428& B8I& yes (strong)& Unknown\\
17411$-$3154 &RAFGL 5379& OH/IR& --& No optical counterpart\\
17423$-$1755 &Hen 3$-$1475& B?& yes (strong)& $3\arcsec\times17\arcsec$ / Bipolar\\
18062$+$2410 &SAO 85766, HD 341617&  B& yes(strong)& Unknown\\
\hline
\end{tabular}
\end{table*}

 For the last ten years our group studied a large number of IRAS
sources with dust temperatures similar to post-AGB stars and 
planetary nebulae ($\sim$100$-$200 K), mainly 
through near-infrared photometry (Garc\'{\i}a-Lario et al. 1990; 1997a; 
Manchado et al. 1989a) and optical spectroscopy (Garc\'{\i}a-Lario et al. 1991; 1994; 1997b; 2001; Riera et al. 1995). 
As a result of this multi-wavelength program we have discovered 
heavily obscured late AGB stars and early post-AGB stars still 
surrounded by thick circumstellar envelopes only detectable in the near 
infrared as well as  post-AGB stars with optically bright 
central stars covering all possible spectral types from M to B in what it
seems to be an evolutionary sequence towards the PN stage, 
together with a small number of new 
extremely young PNe (Garc\'\i a-Lario et al. 1997a; 2002), among them, the
youngest PN known so far: the Stingray Nebula (Bobrowsky et al. 1998).

From this list of newly discovered sources, we selected 
a subsample of 30, which are observable from the 
southern hemisphere. This subsample is also representative of the
various evolutionary stages mentioned above, including very late AGB stars 
with no optical counterparts, post-AGB stars with 
all kind of spectral types and proto-PNe showing a 
variety of morphological structures (from round/elliptical to strongly 
bipolar). 

In Table 1 we list the IRAS sources included in the study.
 In this Table we also indicate some of the 
observational characteristics which are relevant for the 
discussion presented in Sect. 5, like their apparent optical morphology,  
spectral type, or the presence/absence of H$\alpha$ emission in their optical 
spectra (all this information taken from Garc\'\i a-Lario et al. 2002, and 
references therein). 

\begin{table*}[t!]
\caption{Summary table with the results obtained. Quoted values are the 
emission line fluxes or 2$\sigma$ upper limits to the expected emission
line fluxes (in units of $10^{-14}\;{\rm erg}\,{\rm cm}^{-2}\,{\rm s}^{-1}$). 
 The equivalent widths (in units of 10$^{-4}$ $\mu$m) are shown in 
brackets for the features detected in  absorption.}
\begin{tabular}{lcccccccccccc}
\\
   \multicolumn{13}{c}{NIR Spectroscopy}\\
\hline
\multicolumn{1}{l}{IRAS name}&
\multicolumn{2}{c}{H$_2$[$v$=1$\rightarrow$0S(1)]}& 
\multicolumn{2}{c}{Br$\gamma$}&
\multicolumn{2}{c}{H$_2$[$v$=2$\rightarrow$1 S(1)]} &
\multicolumn{2}{c}{CO[$v$=2$\rightarrow$0]}& 
\multicolumn{2}{c}{Pf$\gamma$}& 
\multicolumn{2}{c}{Br$\alpha$}\\ 
\hline
05284+1945  & $\dots$ & & n.d. &$\le$1.3 & $\dots$ & & $\dots$ & & $\dots$ & & $\dots$ & \\
05341+0852  & c & $\le$0.7&  A & [2.2]& $\dots$ & & c & $\le$1.4 & $\dots$ & & $\dots$ & \\
06530$-$0213& c & $\le$1.0&  A & [2.1]& $\dots$ & & $\dots$ & & $\dots$ & & n.d. & $\le$38.9\\
06556+1623  & E & 3.4 & E & 63.2  & c & $\le$2.0 & c & $\le$4.1 & E & 13.0 &$\dots$ & \\
07027$-$7934& E & 2.4 & E & 13.4 & c & $\le$1.9  & c & $\le$4.0 & E & 16.4& E &62.5 \\
07134+1005  & c & $\le$8.0& A & [3.5]& $\dots$ & & $\dots$ & & $\dots$ & &$\dots$ & \\
07331+0021  & c & $\le$39.4 & c & $\le$49.4 & $\dots$ & & $\dots$ & & $\dots$ & & c & $\le$97.4 \\
08005$-$2356 & c & $\le$9.0& c & $\le$13.0& $\dots$ & & E & 24.8 & c &$\le$47.5& c & $\le$21.0 \\
08187$-$1905 & c & $\le$4.2& A & [2.4]& $\dots$ & & $\dots$ & & $\dots$ & & c &$\le$38.5 \\
08242$-$3828 & c & $\le$30.0 & c & $\le$21.4& $\dots$ & & $\dots$ & & $\dots$ && c & $\le$32.2 \\
08425$-$5116 & c & $\le$0.5  & c & $\le$0.5 & $\dots$ & & A & [4.7]& $\dots$ & & c & $\le$17.3 \\
09024$-$5019 & c & $\le$0.4  & E & 18.8]& $\dots$ & & c & $\le$0.3 & E & 12.2& E & 86.0 \\
10178$-$5958 & E & 37.1$^*$& E & 25$^*$& E & 1.4  & E & 2.9 & E & 9.3 & E & 78.4 \\
10197$-$5750 & E & 8.9& E & 22.3& c & $\le$3.4& c & $\le$7.1& E & 5.6 & E & 46.7\\
10215$-$5916 & c & $\le$203.7 & c & $\le$198.3& $\dots$ & & $\dots$ & & $\dots$& & E & 49.1 \\
11438$-$6330 & c & $\le$0.4& $\dots$ & & $\dots$ & & $\dots$ & & c & $\le$27.5 &c & $\le$55.5 \\
12067$-$4508& c & $\le$3.0& c & $\le$4.9& $\dots$ & & $\dots$ & & $\dots$ & & c& $\le$15.9 \\
12175$-$5338 & c & $\le$1.4 & A & [2.7]& $\dots$ & & c & $\le$3.6 & $\dots$& &n.d. & $\le$18.1 \\
13421$-$6125 & c & $\le$0.4 & c & $\le$0.3 &$\dots$ & &$\dots$ & &$\dots$ & &n.d. & $\le$18.1 \\
14104$-$5819 & c & $\le$1.8 & c & $\le$1.2& $\dots$ & & $\dots$ & & $\dots$& &n.d. & $\le$17.8 \\
14331$-$6435 & E & 10.1 & E & 3.4 & E & 1.2& $\dots$ & & $\dots$ & & n.d. & $\le$27.2 \\
16342$-$3814 & c & $\le$5.1 & c & $\le$4.8& $\dots$ & & $\dots$ & & $\dots$ & & n.d. & $\le$15.4 \\
16594$-$4656 & E & 62.7$^*$& E & 9.2& E & 1.9 & c & $\le$2.5 & c & $\le$15.3 & E & 13.3 \\
17119$-$5926 & E & 4.1 & E & 43.7 & E & 2.7& c & $\le$0.4& E & 9.7 & E & 125.5 \\
17149$-$3053 & c & $\le$3.2& c & $\le$3.8& $\dots$ & & $\dots$ & & $\dots$ & &$\dots$ & \\
17150$-$3224 & E & 5.5$^*$& A & [3.1]& c & $\le$0.7& c & $\le$0.9& $\dots$ &&$\dots$ & \\
17311$-$4924 & E & 36.3& E & 2.0& E & 3.2& c & $\le$1.3& n.d. & $\le$8.9& n.d. &$\le$14.1 \\
17411$-$3154 & c & $\le$0.4& c & $\le$0.4& $\dots$ & & $\dots$ & & $\dots$ & & $\dots$ & \\
17423$-$1755 & c & $\le$3.7& E & 26.2& $\dots$ & & E & 31.7& c & $\le$27.0& E &38.1 \\
18062+2410 & E & 3.0& E & 0.8& $\dots$ & & c & $\le$0.5 & c & $\le$14.2& n.d. & $\le$26.7\\
\hline
\end{tabular}
\begin{tabbing}
FIELDxxxxxx\====\=DESCRIPTIONxxxxxxxxxxxxxxxxxxxxxxxxxxxxxxxxxxxx\kill
E\>: emission\\
A\>: absorption\\
c\>: continuum\\
n.d.\>: not detected\\
$\dots$\>: not tried\\
$^*$ \>: mean value of the integrated line fluxes in both runs\\
\end{tabbing}
\end{table*}

\section{Observations and data reduction}

The observations were conducted in  March 1993 and January 1994 with IRSPEC
attached to the Nasmyth focus of the 3.5 m NTT telescope at  La Silla, ESO.
IRSPEC was a cryogenically cooled scanning grating IR spectrometer with a
58$\times$62  pixel array from SBRC as detector (Moorwood et al. 1986). It
covered the  1$-$5 microns wavelength range with a spectral resolution between
1500 and  2000. The spatial scale was 2.2\arcsec/pixel, the slit width was
4.5\arcsec and the un-vignetted field was $\sim$100\arcsec. 
The array had a quantum efficiency of 0.89 at 2.85 $\mu$m, and a well
capacity of  1 10$^{6}e^{-}$. This allowed  exposures up to 60 seconds per
frame in the 2$-$2.6 $\mu$m region.  For integration times around 10 seconds,
the combined dark and read noise was $\sim$150 e$^-$. Owing to the poor
wavelength coverage of the  spectrograph in a single exposure, different
exposures were  needed centered at various wavelengths, each one covering a
different  spectral feature under analysis. The lines observed were the
molecular ro-vibrational lines of molecular hydrogen  $v$=1$\rightarrow$0 S(1)
2.122 $\mu$m and $v$=2$\rightarrow$1 S(1) 2.248 $\mu$m, the recombination 
lines of hydrogen Br$\gamma$ 2.166 $\mu$m,  Pf$\gamma$ 3.741 $\mu$m and 
Br$\alpha$ 4.052 $\mu$m and the CO[$v$=2$\rightarrow$0] 2.294 $\mu$m first 
overtone bandhead. Total on-source integration times were typically of
240 s with exposures ranging from 36 s to 1440 s in the most extreme cases.

The slit orientation was north-south (P.A. $=$ 0\degree) in 1993
with the exception of IRAS 10178$-$5958 (P.A. $=$ 70\degree), 
IRAS 16594$-$4656 (P.A. $=$ 70\degree), IRAS 16342$-$3814 
(P.A. $=$ $-$70\degree) and IRAS 17423$-$1755 (P.A. $=$ $+$55\degree), while
in 1994 the slit orientation was always  east-west (P.A. $=$ 90\degree).
The changes in the slit orientation during the 1993 run were in principle 
made in order to follow the bipolar axis of some well known extended objects, 
like the ones listed above. However, due to a mistake in 
the determination of the sign, IRAS 17423$-$1755 was observed
almost perpendicular to the actual bipolar axis in 1993 while in IRAS
16342$-$3814 independently of the slit orientation, we cover the source
completely. The only other 
known extended bipolar objects in our sample are IRAS 10197$-$5750 and
IRAS 17150$-$3224. IRAS 10197$-$5750 
was only observed in 1994, when the orientation of the slit was 90\degree,
while the bipolar axis of this source is located at P.A. $\sim$ 20\degree.
This means that the slit was positioned also very close to the direction 
perpendicular to its bipolar axis. In the case of IRAS 17150$-$3224, we
observed the source both in 1993 and 1994 with orientations of 0 and 90 
\degree respectively, while the actual bipolar axis of the source is now
known to be located at P.A. $\sim$ $-$45\degree. Two other sources 
(IRAS 10178$-$5958 and IRAS 16594$-$4656) also were observed twice. The implications of these
misadjustments in the analysis of the results obtained 
will be discussed later.

The standard beam$-$switching technique was used in order to subtract the sky
background. The data reduction process includes bias and flat-field
corrections, as well as wavelength calibration, sky subtraction, and absolute
flux calibration using reference standard stars.  For this we made use of the
S/W package IRSPEC installed on MIDAS (version November 1998).

\section{Results}

In Table 2 we list  
the detections/non-detections made as a result of the observations
described in Sect. 3.  

All sources but one
in the sample were observed in the wavelength region surrounding 
the H$_2$[$v$=1$\rightarrow$0 S(1)] line,  resulting in positive detections 
in 10 cases. Out of them, 9 sources were also observed in 
the weaker H$_2$[$v$=2$\rightarrow$1 S(1)] line, among which 
5 showed some emission.

 The Br$\gamma$ line was detected in 17 objects, either in emission (11)
or in absorption (6). In a few cases, observations were also performed
in other spectral lines like Pf$\gamma$, and Br$\alpha$, or around the
CO[$v$=2$\rightarrow$0] first overtone bandhead, 
in order to obtain complementary information. 

The quantitative results are also shown in Table 2. Quoted values are the
emission line fluxes derived by fitting gaussians to the observed line profiles
or 2$\sigma$ upper limits to the expected emission line fluxes in the
case of those sources with no emission detected over the underlying near
infrared continuum. Both are shown in units of  $10^{-14}\;{\rm erg}\,{\rm
cm}^{-2}\,{\rm s}^{-1}$.  We have also estimated the equivalent widths of those
features detected in  absorption (in units of 10$^{-4}$ $\mu$m). The values
obtained are shown in Table 2 within brackets. 

In a few cases, we did not detect any significant signal in the 2D spectra (nor
even a continuum). These sources  have been classified in Table 2 as not
detected. In these latter cases we can generally infer the object's position on
the 2D frames based on information obtained at other spectral lines and
estimate the associated rms of the pixels where the feature investigated 
had been detected. Then we use these rms to calculate the quoted values 
as 2$\sigma$ upper limits to the expected emission line fluxes in Table 2. 

We estimate that for a typical 240 s total on-target exposure and when the
underlying near infrared continuum is weak our 2$\sigma$ detection threshold 
is of the order of $\sim$ $10^{-15}\;{\rm erg}\,{\rm cm}^{-2}\,{\rm s}^{-1}$
for the H$_2$[$v$=1$\rightarrow$0 S(1)], Br$_\gamma$,  
H$_2$[$v$=2$\rightarrow$1 S(1)] and CO[$v$=2$\rightarrow$0] lines and of 
$\sim$ $10^{-14}\;{\rm erg}\,{\rm cm}^{-2}\,{\rm s}^{-1}$ in the Pf$\gamma$ and
Br$\alpha$ regions. Note that the 2$\sigma$ upper limits in Table 2 are
very different from source to source because the rms is dependent of the
continuum. 

Three objects (IRAS 10178$-$5958, IRAS 16594$-$4656 and IRAS 17150$-$3224),
were observed in both runs (1993 and 1994) and for them we quote in Table 2 a
mean value of the integrated line fluxes (the flux differences from epoch to
epoch are less than 20\%). However, the emission detected was always found to be
consistent with a point-like  nature of the observed source with the exception
of IRAS 10178$-$5958 and IRAS 17119$-$5926, where the emission found is
clearly  extended (see Sect. 5.5).

Spectra for all objects (except for IRAS 05284$+$1945, whose  near infrared
counterpart was not detected at any wavelength) in Br$\gamma$,
H$_2$[$v$=1$\rightarrow$0 S(1)], H$_2$[$v$=2$\rightarrow$1 S(1)],
CO[$v$=2$\rightarrow$0], Pf$\gamma$ and Br$\alpha$ are presented in Figures 1a
to 1f. For those sources that were observed twice we show the best spectrum. 

\section{Discussion}

\subsection{Hydrogen recombination lines}

In general, the strength of the near infrared hydrogen lines detected  in 
the stars in our sample is 
consistent with the spectral types derived from optical data. All objects
where a Br$\gamma$ absorption has been detected
show  H$\alpha$ absorption in the optical or
very weak emission, while  those  with Br$\gamma$ strongly in 
emission in the near infrared are also strong H$\alpha$ emitters in the 
optical, as expected.  

In the case of the late AGB stars included in our study
 (those classified as OH/IR stars in Table 1)
the Br$\gamma$ line region is veiled by the strong dust 
continuum emission and we are not able to detect any signature of this
hydrogen recombination line.
These stars are pulsating  with very long periods so heavily obscured by 
their circumstellar envelopes that they usually do not
show any optical counterpart. 

The same veiling effect might be the reason why IRAS 08005$-$2356 
shows a featureless continuum around the Br$\gamma$ line as well as 
at other near infrared hydrogen recombination lines, while in the optical
a relatively strong H$\alpha$ emission is detected. Another possibility is 
that the emission is just strong enough to fill in the photospheric
absorption and, thus, both components compensate each other, leading
to a flat continuum.

 IRAS 09024$-$5019 seems to be also a remarkable case, since it shows a clear 
emission in the light of Br$\gamma$ and Br$\alpha$ while it does not have
any (known) optical counterpart.  This star could be a heavily obscured 
transition object in which the onset of the ionization is now taking place.

\begin{figure*}
\hskip 0cm
\centerline{\psfig{figure=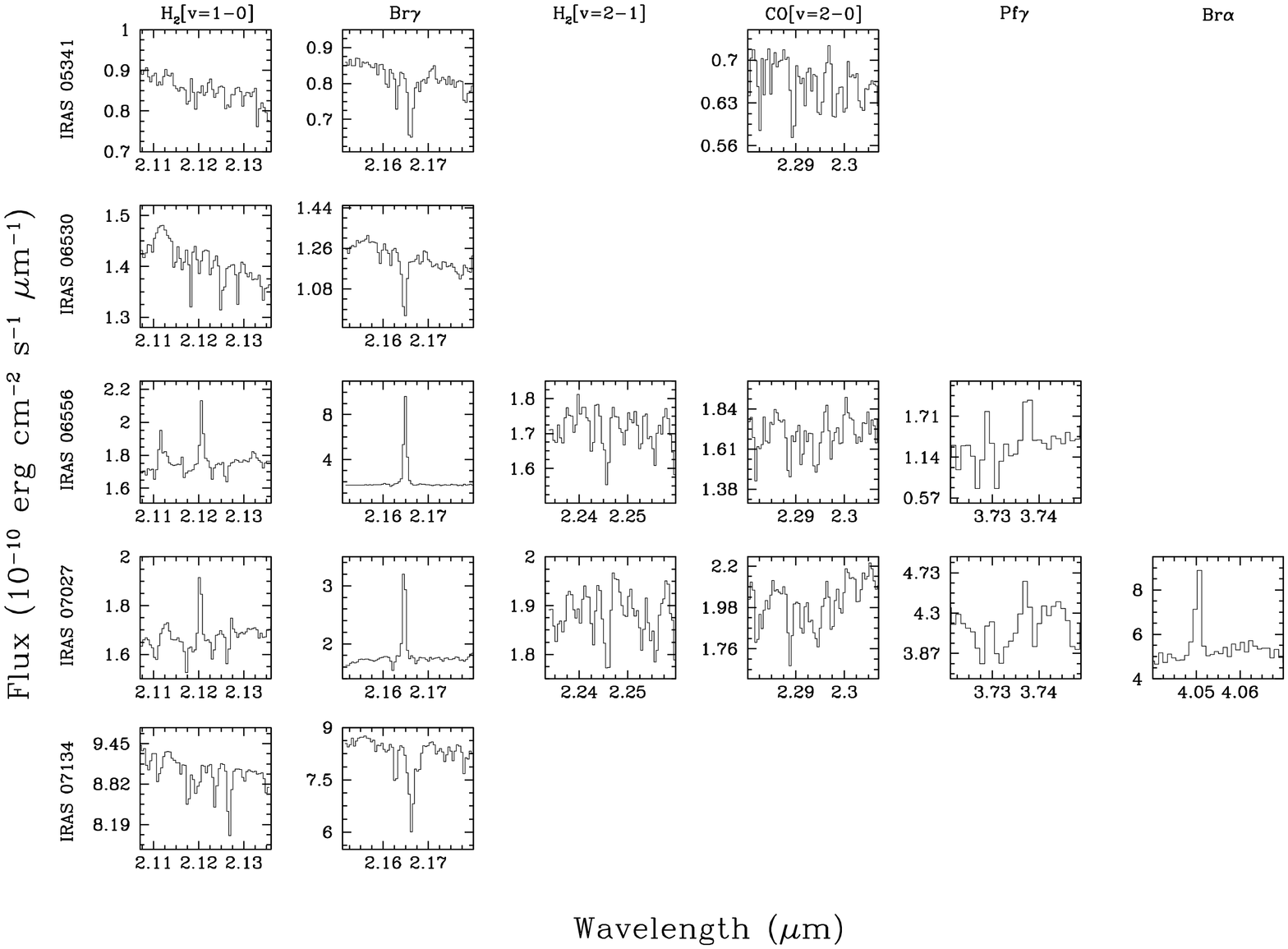,width=15cm,height=18cm,angle=-90}}
\figurename{ 1a. Near infrared spectra of the sources detected with IRSPEC. 
Each row corresponds to observations of a different object. 
The object name is shown on the left along the vertical direction 
and they are displayed in increasing R.A. order}
\end{figure*}

\begin{figure*}
\hskip 0cm
\centerline{\psfig{figure=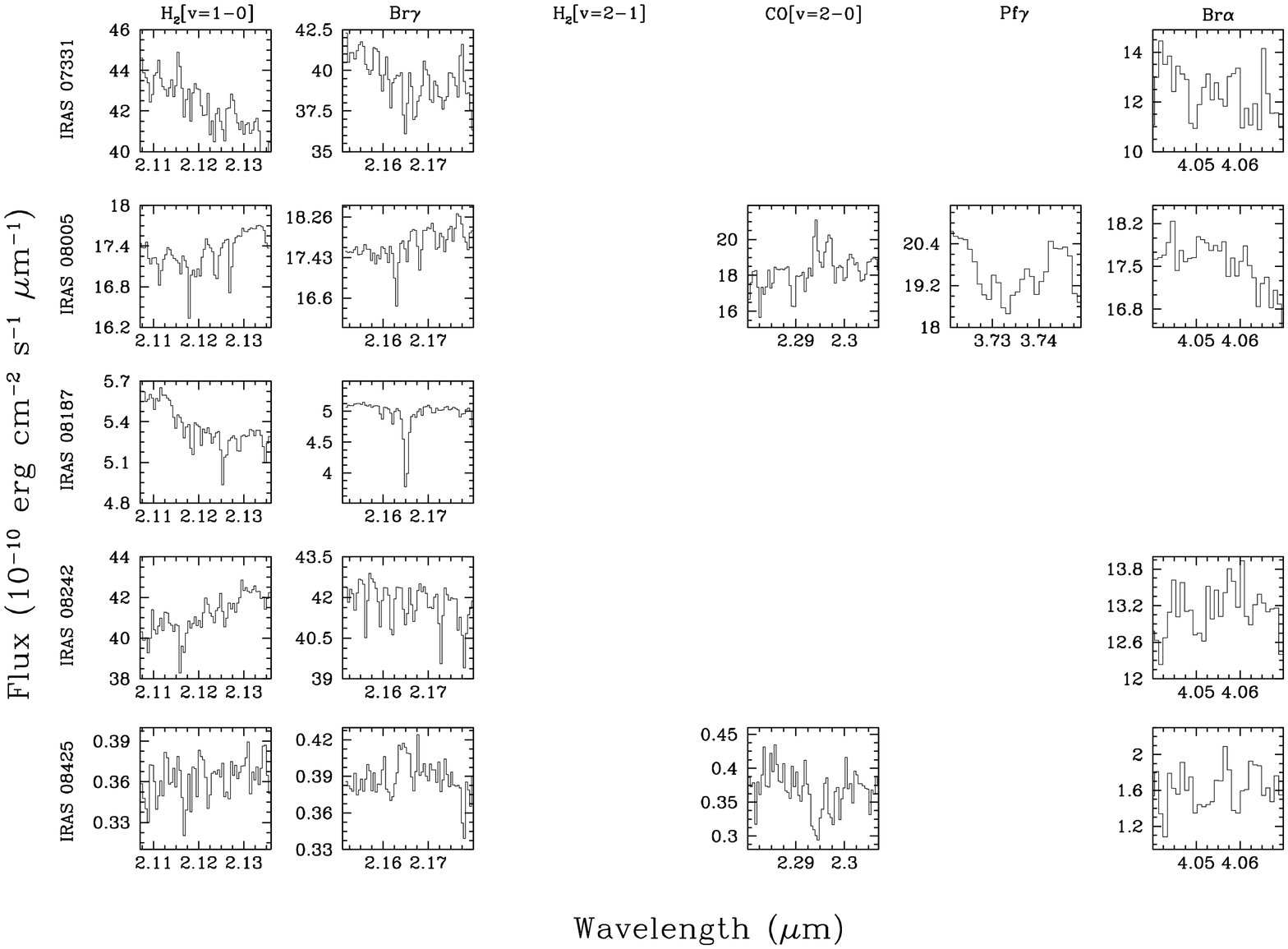,width=15cm,height=18cm,angle=-90}}
\figurename{ 1b. Near infrared spectra of the sources detected with IRSPEC
(cont.)}
\end{figure*}

\begin{figure*}
\hskip 0cm
\centerline{\psfig{figure=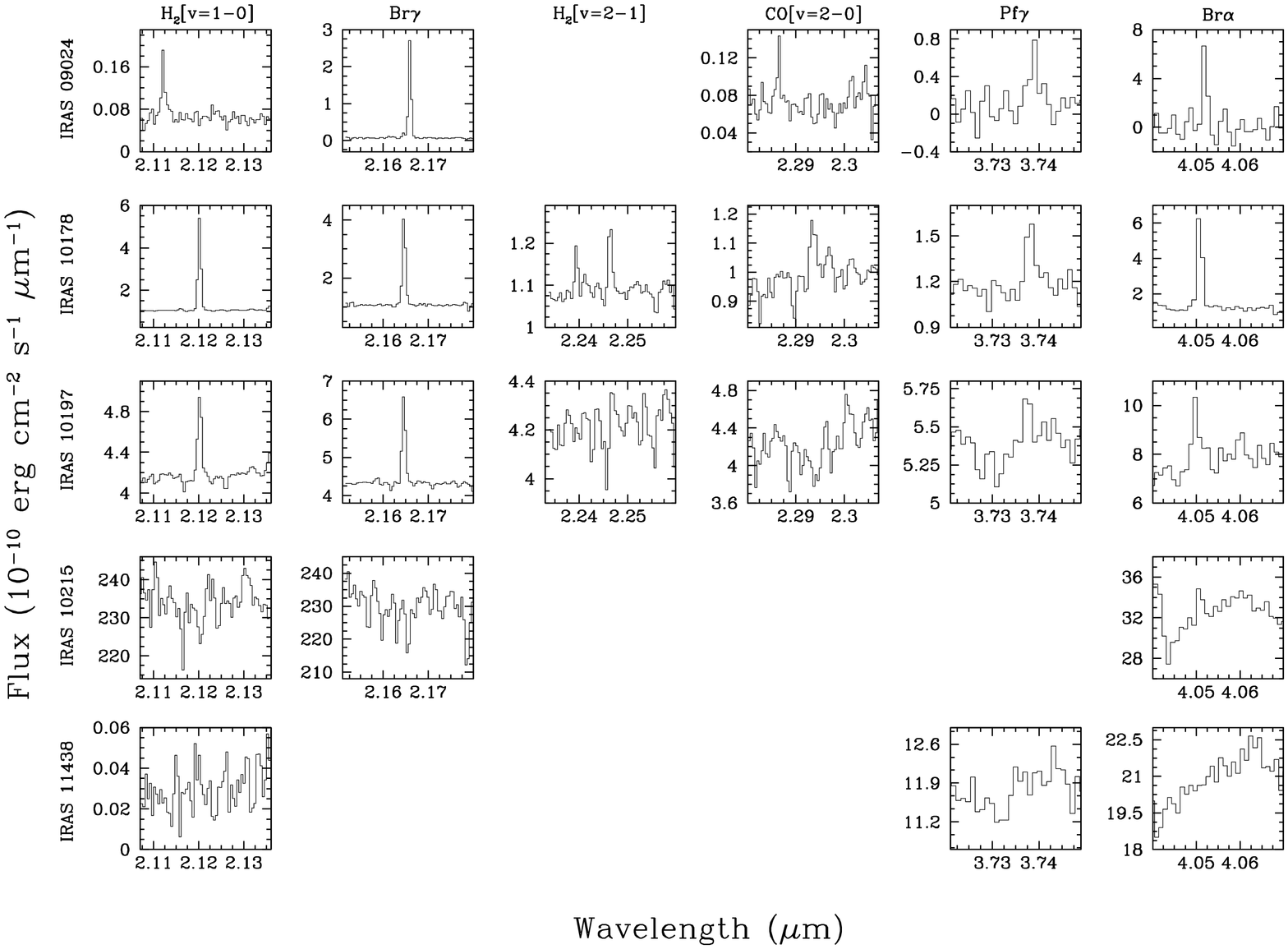,width=15cm,height=18cm,angle=-90}}
\figurename{ 1c. Near infrared spectra of the sources detected with IRSPEC
(cont.). The best H$_2$[$v$=1$\rightarrow$0 S(1)] and Br$\gamma$ spectra for IRAS 10178$-$5958
correspond to the 1994 run}
\end{figure*}

\begin{figure*}
\hskip 0cm
\centerline{\psfig{figure=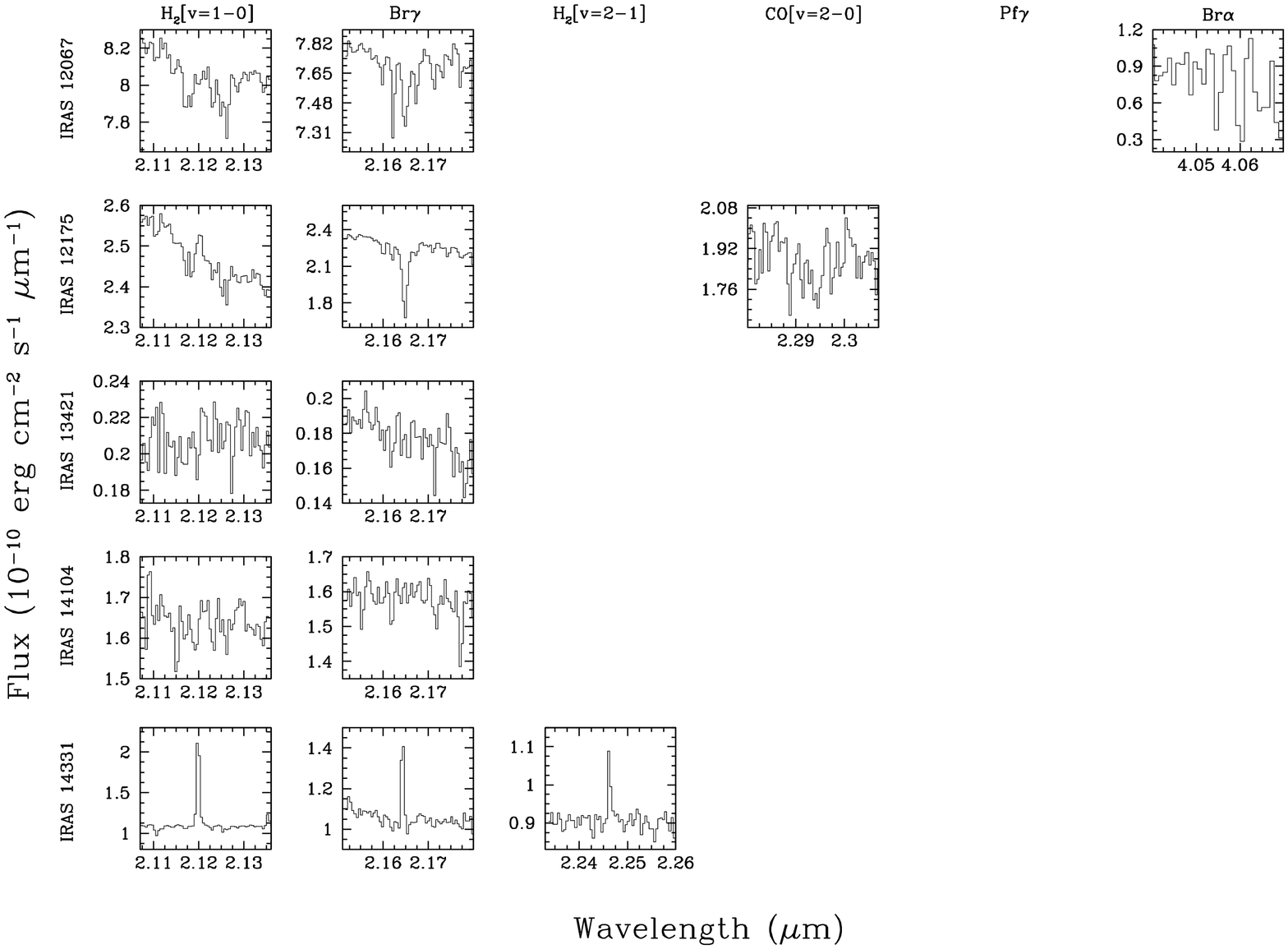,width=15cm,height=18cm,angle=-90}}
\figurename{ 1d. Near infrared spectra of the sources detected with IRSPEC
(cont.)}
\end{figure*}

\begin{figure*}
\hskip 0cm
\centerline{\psfig{figure=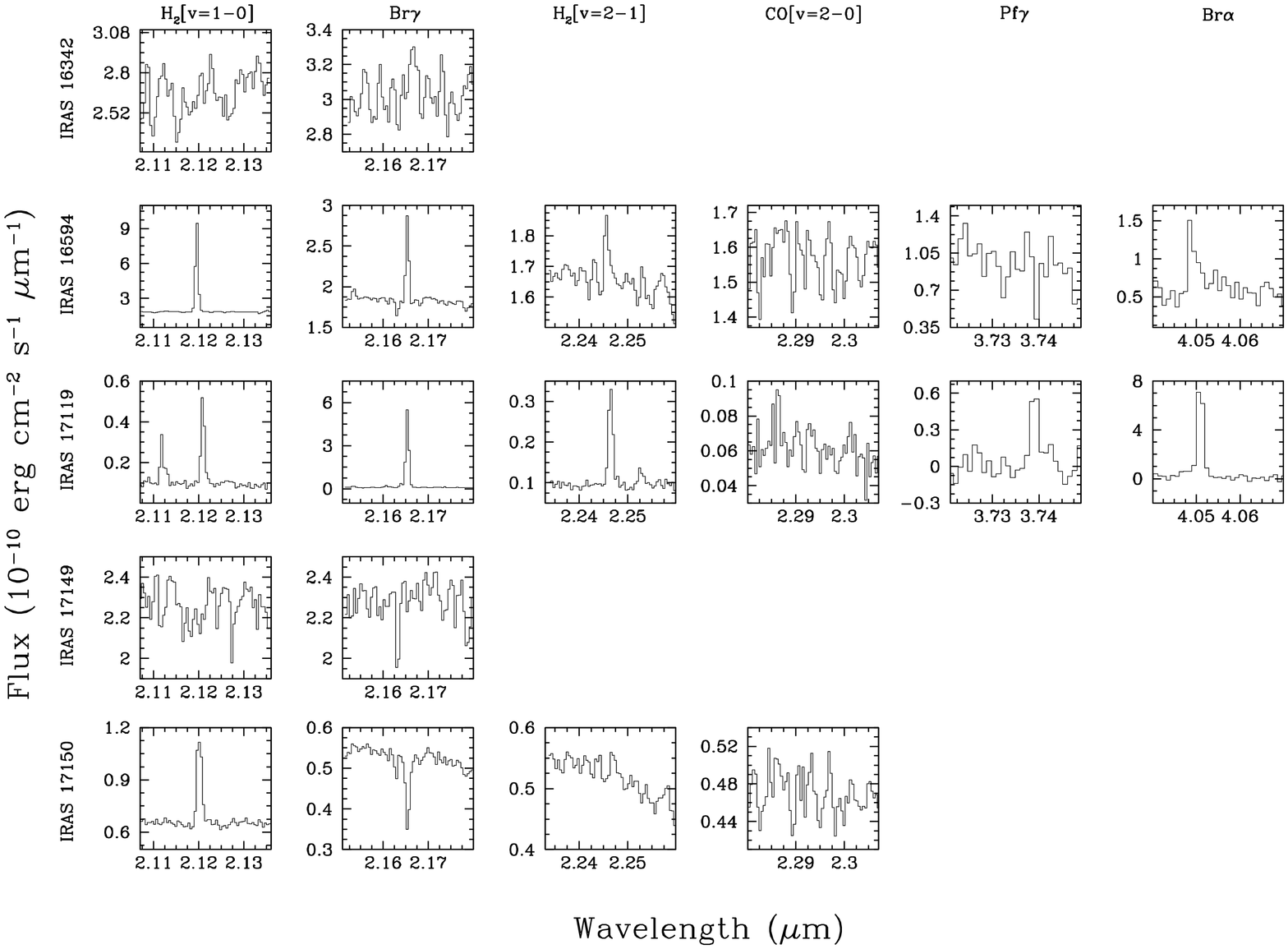,width=15cm,height=18cm,angle=-90}}
\figurename{ 1e. Near infrared spectra of the sources detected with IRSPEC
(cont.). The best H$_2$[$v$=1$\rightarrow$0 S(1)] spectra for IRAS 16594$-$4656 and 
IRAS 17150$-$3224 correspond to the 1994 run}
\end{figure*}

\begin{figure*}
\hskip 0cm
\centerline{\psfig{figure=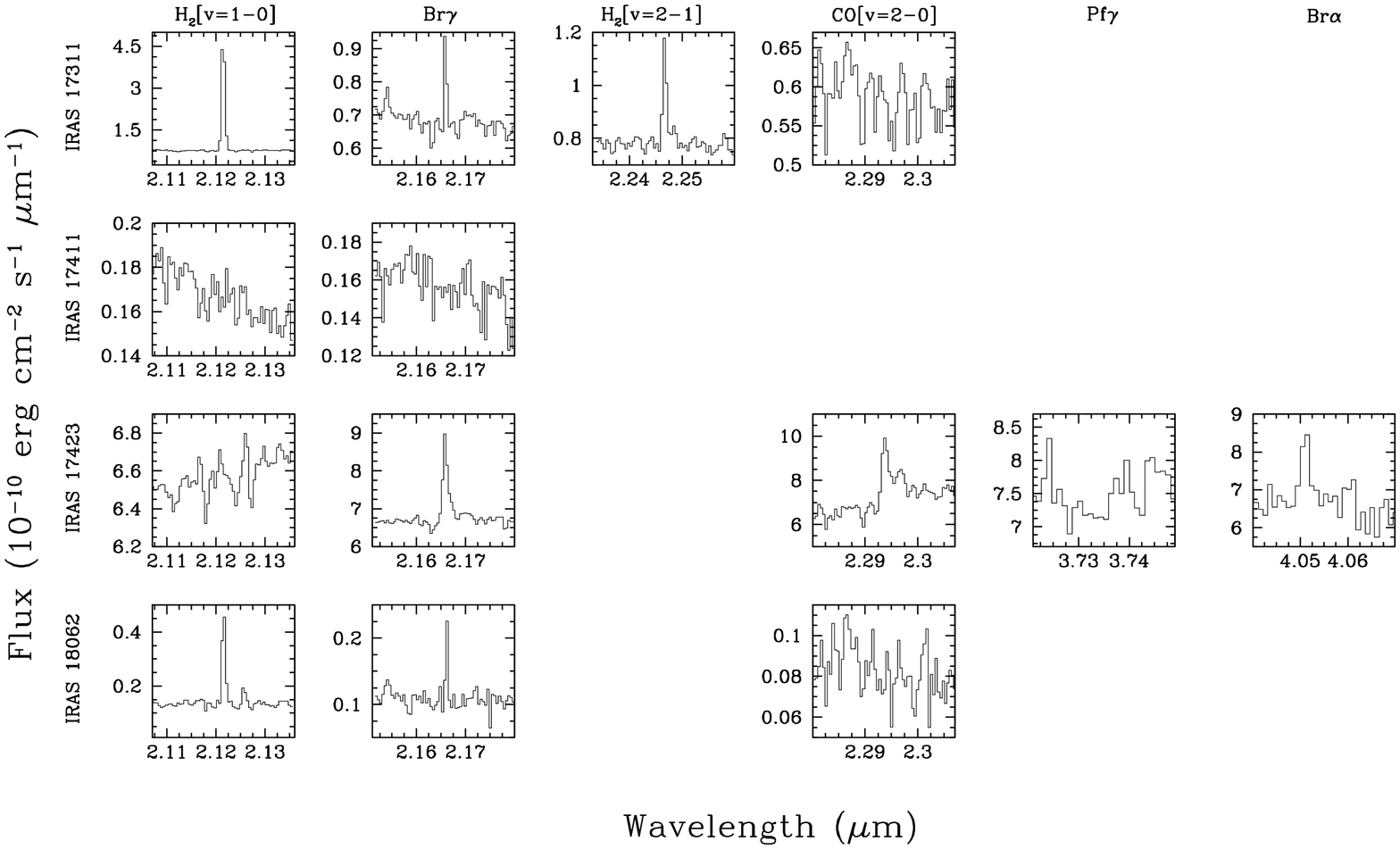,width=15cm,height=18cm,angle=-90}}
\vskip -2cm
\figurename{ 1f. Near infrared spectra of the sources detected with IRSPEC
(cont.)}
\end{figure*}

\subsection{CO first overtone}

 The effective temperature of late AGB stars is expected to be
very low ($\le$2500 K). Consistent with this, a strong 
CO photospheric absorption was detected in IRAS 08425$-$5116 at 2.29 
$\mu$m, the only late AGB star included in
our sample for which we obtained a spectrum at this wavelength.

 This feature corresponds to  the CO first overtone bandhead, which is 
expected to be visible in stars with effective temperatures below 5000 K. 
 However, IRAS 17150$-$3224, a post-AGB star with 
a cool central star (G2I),  did not show any
signature of this bandhead, probably because of the veiling produced by the 
strong thermal emission from the hot dust formed in the 
circumstellar envelope, which completely dominates the observed spectral 
energy distribution at this wavelength. 

  Noteworthy, in three of the post-AGB stars with the earlier spectral types
in the sample we detected the CO first overtone bandhead in emission. This 
has also been observed in a few other transition objects before 
(Oudmaijer et al. 1995) and it has been
interpreted as the result of active post-AGB mass loss, the CO lines
being formed in the dense outflow. 

\subsection{H$_2$ emission and evolutionary stage}
 
 In order to understand the mechanism of excitation of molecular hydrogen   it
is worth to  try to find a possible correlation between the frequency of
positive detections and the evolutionary stage (spectral type of the central
star) of the sources observed.

 Starting from the less evolved objects, we find that none 
of the late-AGB stars included in the sample
show any indication of the presence of H$_2$ emission at 2.122 $\mu$m.
It is important to note however that, as for the hydrogen recombination lines 
above discussed, the spectral  region around this molecular hydrogen
line appears usually completely dominated 
by the thermal emission coming from the hot dust in the envelope, making
the detection of any spectral feature difficult in this kind of objects.

 More  interesting is the fact that, with the only exception of IRAS 
17150$-$3224, which will 
be discussed below, none of the post-AGB stars in the sample with 
spectral types later than A showed any detectable emission in this line.
This includes 6 F-type, 2 G-type, 1 K-type and 1 M-type post-AGB stars.  
It should be noticed that in these optically bright post-AGB stars the 
veiling effect from the circumstellar dust is small and cannot 
justify the non-detections.

 In contrast, all the stars identified as post-AGB candidates in our  list 
with spectral types earlier than A were detected in molecular  hydrogen at
 2.122
$\mu$m. In this case, the only exception was  IRAS 17423$-$1755, whose central
star has tentatively been  classified as A-type or B-type  and it shows no
evidence of  H$_2$ emission at 2.122 $\mu$m. If this emission 
comes mainly from the bipolar lobes, it may have not been covered by our 
slit because it  was positioned by mistake  
perpendicular to the bipolar axis (see Sect. 3). Consistently, we did not detect
any extended emission from this object in any other spectral line 
(nor even in Br$\gamma$ which is detected very strong in emission).
In addition, this is also a quite peculiar,
extended bipolar  proto-PN which shows a very high velocity field and a 
strong near infrared excess
 (Riera et al. 1995; Garc\'{\i}a-Lario et  al. 1997a). Thus,  the
H$_2$ emission may be veiled by the strong underlying continuum and/or
the molecular hydrogen molecules may have been destroyed by the strong 
stellar winds. 

 Out of the 10 objects in which we detected molecular hydrogen
emission, 9 are new detections (only IRAS 17150$-$3224 was 
previously reported by Weintraub et al. 1998). This brings the total 
number of transition stars in which molecular hydrogen emission has been 
detected from 4 to 13. Note that the H$_2$ emission in IRAS 10178$-$5958 was
previously presented by us (Garc\'{\i}a-Lario et al. 1999b).

 The results above are consistent with the non-detection of
molecular hydrogen emission in the bipolar proto-PNe OH 231.8+4.2, IRAS
07131$-$0147 or IRAS 09371+1212 by Weintraub et al. (1998), all of which 
have late-type M central stars. 

However,  IRAS 17150$-$3224 is not the only  post-AGB star with a spectral
type later than A showing H$_2$ emission in the short list of  known 
H$_2$ emitters. There is another discrepant case in the literature, 
the well-known PPN 
CRL 2688 (Sahai et al. 1998), whose central star is classified as F5Iae, 
although it is suspected to be a binary. 

\subsection{H$_2$ emission and morphology}

 It has  been suggested that molecular hydrogen emission is a signpost
for bipolarity in evolved PNe (Kastner et al. 1996). In order to 
explore the validity of this statement also in the precursors of PNe we
can search for a possible correlation between the detection rate of molecular 
hydrogen emission and the morphology of the sources included in our sample.

 For this,  we will restrict our analysis  in the following 
to those sources for
which morphological information exists. This means that we will not consider, 
of course, any of the heavily obscured objects in the list which do not 
show any optical counterpart (all the late-AGB stars plus IRAS 09024$-$5019),
neither those objects classified as of unknown size/morphology in Table 1.
 Note that this class of objects may actually contain a mixture of 
round/elliptical and bipolar sources, as deduced from recent observations
of similar objects made with HST (although they could look point-like sources
in ground-based CCD images) (e.g. Ueta et al. 2000).

 The remaining sources in the sample were classified as round/elliptical or 
bipolar depending on the 
morphological properties derived from ground-based or HST observations (see
again Table 1). 

 If we concentrate our attention now on the ten sources showing H$_2$ 
emission in our sample we can see that morphological information is 
available for 
six of them, out of which five are classified as bipolar 
in Table 1, while only one is considered to be round/elliptical.

 However, among the bipolar objects in our sample (eight in total), there are
three of them which were not detected in H$_2$. This argues against the idea 
of a one to one correlation between H$_2$ emission and bipolarity. 

 In summary, our main conclusion is that although there seems to be an 
association between bipolarity and the presence of H$_2$ emission 
the correlation is not so strong as previously reported by Kastner et al.
(1996), who derived this conclusion from the observations of well evolved PNe.

\subsection{The excitation mechanism of H$_2$}

 Molecular hydrogen can be excited either in shocks or by UV
photons (Burton 1992). In order to investigate which of these two possible 
mechanisms is responsible for the emission observed in the objects included
in our sample it is useful to  compare the relative strength of the
H$_2$[$v$=1$\rightarrow$0 S(1)] and H$_2$[$v$=2$\rightarrow$1 S(1)] 
lines. From the analysis of this diagnostic ratio 
it is possible to determine whether the excitation is produced by 
shocks (Burton, Hollenbach \& Tielens 1992) or by fluorescence through the 
absorption of UV photons coming from the central star (Black \& van Dishoeck 
1987; Sternberg \& Dalgarno 1989). For moderate UV radiation fields and gas 
densities lower than 10$^{5}$ cm$^{-3}$, shock$-$excited emission is 
expected to be associated to flux ratios H$_2$[$v$=1$\rightarrow$0 
S(1)/$v$=2$\rightarrow$1 S(1)] $\sim$ 10 while in the case of
emission induced by fluorescence a value of $\sim$ 2 (Aspin et al. 1993;
Ramsay 1993; Burton et al. 1998) is predicted. In
Table 3 we show the ratios derived for 9 
out of the 10 post-AGB stars in the sample
detected in molecular hydrogen for which data is available. The estimated
uncertainty is $\sim$20\%. For those objects in which the 
 H$_2$[$v$=2$\rightarrow$1 S(1)] line was not detected lower limits are
given. 

 As we can see in Table 3 a wide range of values is obtained.
If the above mentioned models are applied to the stars in our list
we deduce that shock$-$excited H$_2$ emission must be the dominant excitation 
mechanism in at least three of the objects in our sample with a 
clearly marked bipolar morphology: 
IRAS 10178$-$5958 , IRAS 16594$-$46456 (both of 
uncertain spectral type) and IRAS 17150$-$3224 (G-type central star) 
as well as in two other objects of unknown morphology: 
IRAS 14331$-$6435 and IRAS 17311$-$4924 (both with B-type central stars),
while fluorescence excitation seems to be the prevalent excitation mechanism
in the Stingray Nebula (IRAS 17119$-$5926), which is a very young PN showing
an  incipient bipolar morphology. For the remaining objects 
the lower limits found are ambiguous to support either fluorescence or 
shock excitation. 
 
 From the results obtained, there seems to be a prevalence of 
shocked-excitation in those objects showing the most extreme 
bipolar morphologies. However, the results are not conclusive since  
the statistics are very poor. There is also no evident
connection between the excitation mechanism derived from the analysis of the 
H$_2$[$v$=1$\rightarrow$0 S(1)/$v$=2$\rightarrow$1 S(1)] ratio
and the spectral type of the central star in which molecular hydrogen has
been detected.  However, in this latter analysis we must be careful since 
it is well known that episodic mass loss 
during the post-AGB phase can induce significant spectral changes which 
may alter for some time the observable spectral type and, thus, taking 
the spectral type as an indicator of the evolutionary stage might be not
the best choice (Bobrowsky et al. 1998; Garc\'\i a-Lario et al. 2001).

 In addition,  an important caveat needs to be considered 
when interpreting the data in Table 3. This is the fact
that the models above mentioned lose their validity at very high densities 
($n_e$ $>$ 10$^{5}$ cm$^{-3}$) and we know that some of the post-AGB stars 
and  transition objects under study are surrounded by regions of these
very high densities.  In this case,  
collisional de-excitation of fluorescence excited levels (`collisional 
fluorescence') might become important, increasing the $v$=0 and 1 populations 
over the pure fluorescent values, leading to higher 
H$_2$ [$v$=1$\rightarrow$0 S(1)/$v$=2$\rightarrow$1 S(1)] 
ratios (Sternberg \& Dalgarno 1989; Burton et al. 1990) which could
mimic shock-excitation.

 In order to address the problem from a different point of view it is also
useful to study the detailed spatial distribution of the molecular hydrogen 
emission and compare this, whenever possible with the spatial 
distribution  observed in the light of Br$\gamma$. 

 The H$_2$[$v$=1$\rightarrow$0 S(1)]/Br$\gamma$ ratio
has recently  found to be correlated both with the evolutionary stage and
the degree of bipolarity in a 
sample of well known bipolar PNe (Guerrero et al. 2000). Bipolar PNe with
small size, poorly
defined rings and bright central stars exhibit  small 
H$_2$[$v$=1$\rightarrow$0 S(1)]/Br$\gamma$ ratios ($\sim$ 0.1$-$0.5)
 while those with  large, well defined 
rings and faint central stars showed larger
H$_2$[$v$=1$\rightarrow$0 S(1)]/Br$\gamma$ ratios ($>$5). 

 Guerrero et al. (2000) suggest that the excitation of  the molecular hydrogen
in each of these groups 
turns from fluorescence excitation at the early stages of PNe
to shocked-excited emission as the bipolar PNe evolve. This is supported by 
the different spatial distribution of the observed molecular hydrogen 
emission.

 H$_2$-dominated bipolar PNe are found to show different morphological
and physical properties. Their central regions have low surface brightness, 
in agreement with their low densities while Br$\gamma$-dominated bipolar 
PNe have a high surface brightness and larger densities. 

 If this effect  is interpreted as an evolutionary sequence (Balick 1987)
then the less evolved bipolar PNe with molecular hydrogen emission would 
start their evolution as Br$\gamma$ dominated becoming later H$_2$-dominated. 

 Note, however, that this scenario does not apply to all PNe. 
Well evolved, low-mass density-bounded round/elliptical PNe 
are not expected to preserve hydrogen in the form of molecules since most 
(if not all) of the hydrogen in the circumstellar envelope becomes ionized
as the ionization front passes through the gas. These PNe will end
their lives as Br$\gamma$-dominated, unlike the  group of
PNe with more massive progenitors. Those with more massive progenitors 
are expected to develop ionization-bounded nebulae, where a substantial 
fraction of the mass ejected will never become ionized. This enables 
the detection of neutral material in the circumstellar gas even 
at a very late stage as PNe, provided the UV radiation field from the
central star is hard enough to produce fluorescence excitation. 
In the subgroup of bipolar PNe, the 
shielding provided by the dense equatorial regions that collimate the outflow 
from the UV photons escaping from the central star would also favor the
preservation of hydrogen in molecular state. 
This would explain the high detection rate 
of molecular hydrogen emission in bipolar type I PNe, which are suspected 
to be the result of the evolution of stars with large progenitor masses.
In this latter case, molecular hydrogen can also be excited by shocks
formed in the high velocity outflow. 

Applying the above considerations to the analysis of the molecular 
hydrogen emission observed in precursors of PNe
it would be natural to expect fluorescence to be activated when the 
radiation coming from the central post-AGB star becomes hard enough to 
produce a significant number of UV photons  as the central 
star moves in the HR diagram
towards hotter effective temperatures in its way to becoming a PN.
 On the other hand, the interaction of the fast post-AGB wind 
with the slow-moving material ejected during the previous AGB phase
would favor the formation of shocks, especially in those objects showing a 
strong collimation induced by the presence of a thick equatorial disk or 
torus, in the process of developing a bipolar morphology. 

\begin{table*}[t!]
\caption{Flux ratios observed in the sources showing molecular hydrogen 
emission and prevalent excitation mechanism tentatively inferred from the data}
\begin{tabular}{lcccl}\hline 
Object& H$_2$[$v$=1$\rightarrow$0 S(1)]$^1$& H$_2$[$v$=1$\rightarrow$0
S(1)]/H$_2$[$v$=2$\rightarrow$1 S(1)]& 
H$_2$[$v$=1$\rightarrow$0 S(1)]/Br$\gamma$& Excitation \\
\hline
IRAS 06556+1623 &  3.4&  $>$1.7&    0.05& Fluorescence\\ 
IRAS 07027-7934 &  2.4&  $>$1.3&    0.18& Fluorescence\\ 
IRAS 10178-5958 & 37.1&      26.5&    1.50& Shock\\
IRAS 10197-5750 &  8.9&  $>$2.6&    0.40& Fluorescence$^{*}$\\
IRAS 14331-6435 & 10.1&     8.4&    2.97& Shock\\
IRAS 16594-4656 & 62.7&      33&    6.82& Shock\\
IRAS 17119-5926 &  4.1&     1.5&    0.09& Fluorescence\\ 
IRAS 17150-3224 &  5.5&  $>$7.8&   ---$^2$& Shock\\
IRAS 17311-4924 & 36.3&    11.3&    18.2& Shock\\
IRAS 18062+2410 &  3.0&     ---$^3$& 3.75& Shock?\\
\hline
\end{tabular}
\\
$^1$Emission line flux in units of 10$^{-14}$ erg cm$^{-2}$ s$^{-1}$\\
$^2$Br$\gamma$ detected in absorption\\
$^3$Not observed in the H$_2$[$v$=2$\rightarrow$1 S(1)] line\\
$^{*}$ Shock-excited emission can be present in the outer lobes as deduced
from recent HST images but they were not covered by our slit
\end{table*}

 In order to check whether this scenario is consistent with the data 
obtained on the stars in our sample we have derived 
the H$_2$[$v$=1$\rightarrow$0 S(1)]/Br$\gamma$ ratios 
for the objects that exhibit H$_2$ emission. 
The results obtained are presented in Table 3, where we can see 
that, as for the H$_2$[$v$=1$\rightarrow$0 S(1)/$v$=2$\rightarrow$1 S(1)] ratio,
 a wide range of values is found 
(the quoted values, not corrected for extinction, again have an estimated 
uncertainty of $\sim$20\%). 
 
  Only four sources (out of ten) are found to be Br$\gamma$-dominated.  These
are: IRAS 06556+1623, IRAS 07027$-$7934, IRAS 10197$-$5750 and IRAS
17119$-$5926. Detailed morphological information exists for the latter three
sources.  IRAS 07027$-$7934 is known to be a very young PN showing a
round/elliptical  shape. The bipolar morphology is only incipient in IRAS
17119$-$5926 which otherwise would look also quite round/elliptical. The only
source in this short list with a marked bipolar morphology is IRAS 10197$-$5750
(Sahai et al. 1999), which is seen almost edge-on. This source  has recently
been imaged  with HST  (NICMOS) in the light of  H$_2$[$v$=1$\rightarrow$0
S(1)] (Sahai et al. 2000) and the spatial  distribution of the emission in this
line  is found to be very similar to the distribution of scattered light in the
bright lobes, suggesting that fluorescence excitation is the main mechanism
contributing to the overall observed emission. However, a clear enhancement of
emission is observed at two localized point-symmetric regions  distant from the
central star which unfortunately have not been covered by our slit (the slit
orientation was 90\degree and its bipolar axis is orientated $\sim$
$+$20\degree), which we interpret as  shock-excited emission probably
originated by the interaction  of the high velocity post-AGB wind with the slow
moving old AGB shell. Interestingly, the four Br$\gamma$-dominated sources
show  H$_2$[$v$=1$\rightarrow$0 S(1)/$v$=2$\rightarrow$1 S(1)] ratios
compatible with fluorescence excitation (see Table 3).

 Two other sources in our list have also been imaged in the light of 
H$_2$[$v$=1$\rightarrow$0 S(1)] with HST (NICMOS).    
These are IRAS 10178$-$5958 (Sahai et al. 2000) and IRAS 17150$-$3224 (Su et 
al. 2000) belonging to the group of H$_2$-dominated sources in our
sample.

In both cases, the spatial distribution of the molecular hydrogen emission
suggests a shock-excited origin.  In IRAS
10178$-$5958, the molecular hydrogen emission comes mainly from 
the walls of the bipolar lobes (Garc\'\i a-Lario et al. 1999b), 
while in IRAS 17150$-$3224 the bulk of the detected emission arises in an
expanding molecular torus surrounding the central star (Weintraub et al. 
1998) although several clumps of molecular hydrogen emission  are
also detected in the 
outer regions of the bipolar lobes according to Su et al. (2000). 
A similar spatial distribution was also found by Sahai et al. (1998) in the 
case of AFGL 2688, where the emission observed was also interpreted as due
to shock-excitation.

Unfortunately, none of the four other H$_2$-dominated  sources in Table 3 
have been imaged in molecular hydrogen with HST. 
Three of them (namely IRAS 14331$-$6435, IRAS 17311$-$4924 and 
IRAS 18062+2419) are sources not spatially resolved from ground-based 
observatories. 
Only IRAS 16594$-$4656 was found to be extended from the
ground in optical CCD images and it was later imaged with HST in the 
optical showing an extended, complex multi-polar morphology 
(Garc\'\i a-Lario et al. 1999a; Hrivnak et al. 1999). 
All of them show large H$_2$[$v$=1$\rightarrow$0 S(1)/$v$=2$\rightarrow$1 S(1)] ratios
 which are consistent with shock excitation. 
 
The emission seems to be extended only in the case of IRAS 10178$-$5958 
and IRAS 17119$-$5926 but we did not find any spatial dependence of the  
measured H$_2$[$v$=1$\rightarrow$0 S(1)/$v$=2$\rightarrow$1 S(1)] ratios along
the nebulae. This  confirms shock-excitation as the prevalent
mechanism of excitation of the molecular
hydrogen in IRAS 10178$-$5958 throughout its whole nebular extension.
 On the other hand, fluorescence-excitation seems to be the
mechanism through which molecular hydrogen gets excited 
in the case of IRAS 17119$-$5926. 
  For IRAS 17119$-$5926, 2D spectra of both H$_2$ lines were taken with
different (perpendicular) slit orientations in 1993 and 1994. Remarkably, 
the molecular hydrogen emission seems to be quite homogeneous independent of 
the slit orientation considered 
and extends well beyond the limits of the Br$\gamma$ emission. This suggests 
that most of the emitting molecular hydrogen is located outside the small
ionized region of this very young PN, in the still neutral outer shell.

In summary, when we integrated all the
flux received in each line, we obtain high H$_2$[$v$=1$\rightarrow$0
S(1)/$v$=2$\rightarrow$1 S(1)] and  H$_2$[$v$=1$\rightarrow$0 S(1)]/Br$\gamma$
ratios that are consistent with the prevalence of shock excitation, while
low ratios indicate fluorescence as the dominant mechanism for
 the observed H$_2$ 
emission. This is supported by the different spatial distribution seen
in those objects with existing H$_2$ NICMOS images.

With our small sample and the rest of available data we find different types of
sources that exhibit H$_2$ emission during the post-AGB phase. We propose that
this scenario could be summarized as follows: 

- In some cases, especially strongly bipolar sources with known early spectral
types, there must be a contribution from the two excitation mechanisms in the
observed emission (e.g. IRAS 10197$-$5750). It is very difficult to distinguish
the contribution of both excitation mechanisms but it seems clear that the 
shocked H$_2$ emission is more intense than the fluorescence-excited
emission, leading to higher H$_2$[$v$=1$\rightarrow$0 S(1)/$v$=2$\rightarrow$1
S(1)] and  H$_2$[$v$=1$\rightarrow$0 S(1)]/Br$\gamma$ ratios. 

- The H$_2$ emission that comes from bipolar sources with later than A 
spectral types (e.g. AFGL 2688 and IRAS 17150$-$3224) is shock-excited because the central post-AGB star is not hot
enough to produce fluorescence emission,
which, so it has been demonstrated, is correlated with the evolutionary stage
(spectral type) of the central post-AGB star. So the presence of shock-excited
emission early in the post-AGB phase must be indicative of the evolutionary
morphological stage of the nebulosity that surrounds the central post-AGB star.
However, it remains unclear if the early development of marked bipolar
morphologies during the post-AGB phase could be dependent on the mass of the
central progenitor star. For higher masses, the morphological evolution could
be faster and the formation of bipolar morphologies is easier; so, therefore,
is the detection of shock-excited H$_2$ emission. 

- On the other hand, the fact that H$_2$ emission is detected towards well
evolved bipolar PNe could be indicative that the destruction of neutral
material around the non-bipolar ones (e.g. IRAS 07027$-$7934 or even IRAS 
17119$-$3224) can roughly coincide with the photoionization of the
circumstellar shell. This could  indicate that the fluorescence-excited H$_2$
emission detected in non-bipolar sources  is observed only during a short
period of time in the post-AGB phase. However, in those bipolar objects where
the wind is strongly collimated  by the presence of an optically thick
equatorial disc or torus, the molecular material can be shielded from the
ionizing photons coming from the central star. 

This entire hypothesis deserves further study to understand better the
physical  and chemical processes involved during the short transition phase
from the AGB phase  to the PN stage. At present, the number of known transition
objects and the  obtaining of high quality data are growing continuously. A
detailed spatially resolved spectroscopic analysis of the H$_2$ emission line
spectrum in a large sample of planetary nebulae precursors is needed to 
address these issues fully.

\section{Conclusions}

With the data here presented we have extended the number of post-AGB stars
and/or proto-PNe where molecular hydrogen emission has been detected 
from four to thirteen. 

 Our results confirm the previous finding by Weintraub et al. (1998) made on
a smaller sample of sources that 
the onset of emission from molecular hydrogen takes place in the post-AGB 
phase. In many cases
this occurs well after the generation of the bipolar structure takes place 
and before the nebular envelope gets ionized. 

 In the past it has been suggested that molecular hydrogen emission is a 
signpost for bipolarity in PNe  (Kastner et al. 1996).
We find that the detection of molecular hydrogen
strongly depends on the excitation mechanism.

 The detection of fluorescence-excited molecular hydrogen
emission in precursors of PNe is strongly correlated with the 
evolutionary stage of the central star  (spectral type),
independent of the morphology of the source considered.
The fluorescence excitation of molecular hydrogen emission is produced 
as a consequence of the absorption of the increasing number of UV photons 
escaping from the relatively hot, rapidly evolving central post-AGB star
 by the hydrogen
molecules present in the circumstellar shell. As a general trend, we find
that  fluorescent H$_2$ emission  becomes 
active when the central post-AGB star reaches a certain temperature that
corresponds to an A-spectral type.

 On the other hand, shock-excited emission is only detected in objects with a
marked bipolar morphology, sometimes at a very early stage during the 
post-AGB phase, when the star is still very cool and, sometimes, localized
in the waist and/or in the specific regions of the bipolar lobes (either at
the walls, like in IRAS 10178$-$5958, or at the tips, like in 
IRAS 10197$-$5750), 
where a physical interaction exists between fast- and slow-moving material.

Bipolarity also helps preserving the molecular hydrogen from destruction, 
favoring its detection in well evolved bipolar PNe. Actually, with very 
few exceptions, the brightest H$_2$ emission, shock-excited, is detected
toward the waist of bipolar PNe (Kastner et al. 1996). 
The high correlation found between bipolarity and molecular
hydrogen emission seems to be just the consequence of the bias induced by this
fact. 

The diagnostic ratios H$_2$[$v$=1$\rightarrow$0 S(1)/$v$=2$\rightarrow$1 S(1)]
and H$_2$[$v$=1$\rightarrow$0 S(1)]/Br$\gamma$  are shown to be very good
indicators of the prevalent mechanism of  excitation in proto-PNe.
Unfortunately, the information on the relative spatial distribution of the
different molecular hydrogen lines with  respect to the hydrogen recombination
lines is still sparse. Without this information we cannot determine the precise
localization of  shock-excited and fluorescence-excited regions on individual
sources.  As we have shown through this paper this is a strong limitation if we
want  to understand the physical processes which are taking place in this
still  poorly known transition phase which precedes the formation of PNe. This
is especially important to interpret the emission observed in complex objects
like IRAS 10197$-$5750, where both fluorescence- and  shock-excited regions may
coexist. High spatial resolution images (e.g. as in the past with NICMOS
on board HST) in the three emission lines above mentioned for a statistically 
significant number of transition objects and young PNe will provide this
crucial information. This would be an ideal way to test the scenario here
proposed.  

\acknowledgements
PGL and AM acknowledges support from grant PB97$-$1435$-$C02$-$02 from the Spanish
Direcci\'on General de Ense\~nanza Superior (DGES).

\end{document}